\documentclass[sigconf]{acmart}%
\usepackage{booktabs}
\usepackage{multirow}
\usepackage{graphicx}
\usepackage{subcaption}
\usepackage[table]{xcolor}
\usepackage{tabularx}
\usepackage{listings}
\usepackage{xcolor}
\AtBeginDocument{%
  }

\setcopyright{acmlicensed}
\copyrightyear{2018}
\acmYear{2018}
\acmDOI{XXXXXXX.XXXXXXX}
\acmConference[MM '26]{Proceedings of the 34th ACM International Conference on Multimedia}{November 10--14, 2026}{Rio de Janeiro, Brazil}
\acmISBN{978-1-4503-XXXX-X/2018/06}

\acmSubmissionID{xxx}

\begin{document}

\title{MemCoT: Test-Time Scaling through Memory-Driven Chain-of-Thought}

\author{Haodong Lei}
\email{leihaodong@seu.edu.cn}
\affiliation{%
  \institution{Southeast University}
  \city{Nanjing}
  \state{Jiangsu}
  \country{China}
  }
\affiliation{%
  \institution{Shanghai Artificial Intelligence Laboratory}
  \city{Shanghai}
  \country{China}
  }

\author{Junming Liu}
\email{liujunming@pjlab.org.cn}
\affiliation{%
  \institution{Shanghai Artificial Intelligence Laboratory}
  \city{Shanghai}
  \country{China}
  }

\author{Yirong Chen}
\email{chenyirong@pjlab.org.cn}
\affiliation{%
  \institution{Shanghai Artificial Intelligence Laboratory}
  \city{Shanghai}
  \country{China}
  }

\author{Ding Wang}
\authornotemark[2]
\email{wangding@pjlab.org.cn}
\affiliation{%
  \institution{Shanghai Artificial Intelligence Laboratory}
  \city{Shanghai}
  \country{China}
  }

\author{Hongsong Wang}
\authornotemark[2]
\thanks{\textdagger Ding Wang and Hongsong Wang are corresponding authors.}
\email{hongsongwang@seu.edu.cn}
\affiliation{%
  \institution{Southeast University}
  \city{Nanjing}
  \state{Jiangsu}
  \country{China}
  }

\renewcommand{\shortauthors}{Trovato et al.}

\begin{abstract}
Large Language Models (LLMs) still suffer from severe hallucinations and catastrophic forgetting during causal reasoning over massive, fragmented long contexts. Existing memory mechanisms typically treat retrieval as a static, single-step ``passive matching'' process, leading to severe semantic dilution and contextual fragmentation. To overcome these fundamental bottlenecks, we propose \textbf{MemCoT}, a test-time memory scaling framework that redefines the reasoning process by transforming long-context reasoning into an iterative, stateful information search. MemCoT introduces a multi-view long-term memory perception module that enables Zoom-In evidence localization and Zoom-Out contextual expansion, allowing the model to first identify where relevant evidence resides and then reconstruct the surrounding causal structure necessary for reasoning. In addition, MemCoT employs a task-conditioned dual short-term memory system composed of semantic state memory and episodic trajectory memory. This short-term memory records historical search decisions and dynamically guides query decomposition and pruning across iterations. Empirical evaluations demonstrate that MemCoT establishes a state-of-the-art performance. Empowered by MemCoT, several open- and closed-source models achieve SOTA performance on the LoCoMo benchmark and LongMemEval-S benchmark. Our code is available at \url{https://github.com/Haodong-Lei-Ray/MemCoT}.
\end{abstract}


\begin{teaserfigure}
  \includegraphics[width=\textwidth]{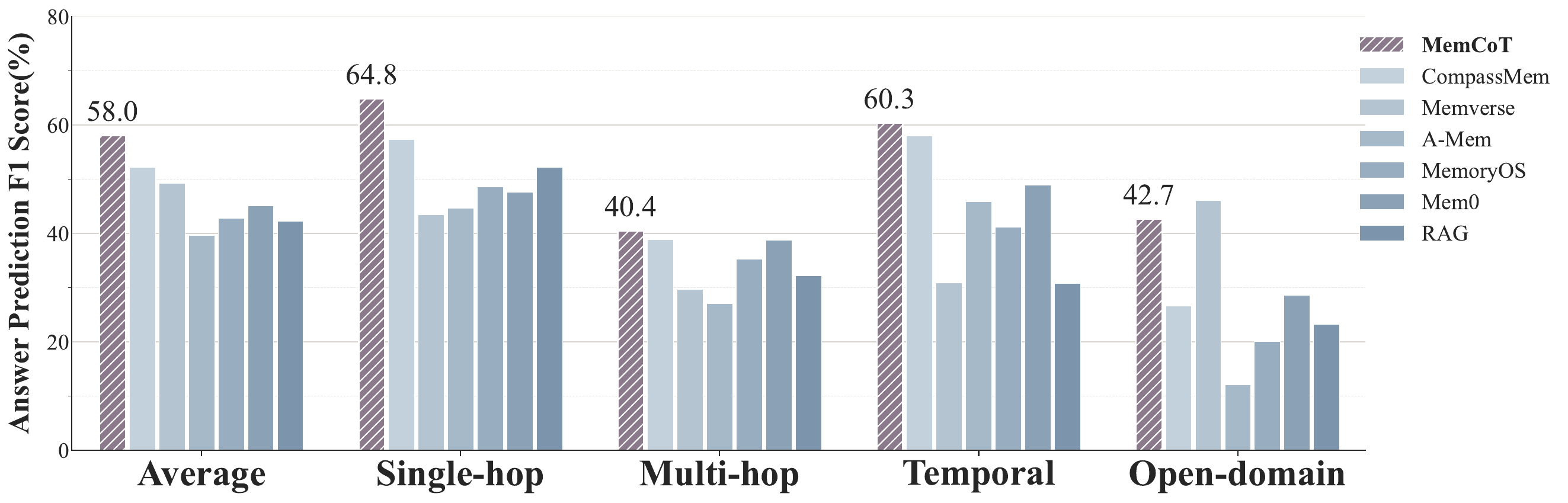}
  \caption{The answer prediction $F_1$ scores (\%) of GPT-4o-mini on the LoCoMo benchmark. \textbf{MemCoT} achieves a state-of-the-art overall $F_1$ score of 58.03\%, significantly outperforming all baseline memory systems across all sub-tasks.}
  \label{fig:Intro}
\end{teaserfigure}


\maketitle

\section{Introduction}
\label{sec:intro}
Despite the rapid expansion of context windows to the million-token scale, Large Language Models (LLMs) still fail remarkably in long-horizon reasoning over extended interaction histories~\cite{bai-etal-2024-longbench, LBT, MemAgent}. In real-world scenarios, models frequently forget earlier user preferences, lose track of causal dependencies across distant dialogue turns, and produce hallucinated conclusions when reasoning requires synthesizing scattered evidence over time~\cite{InfMem}. This paradox reveals a deeper limitation: long-context failure is not primarily caused by insufficient context capacity, but by the model’s inability to perform coherent reasoning over temporally distributed information~\cite{hu2026memoryageaiagents}. Simply enlarging the context window does not resolve this issue; instead, it often amplifies semantic noise~\cite{TPON}, dilutes critical signals, and exacerbates reasoning instability. These phenomena suggest that the challenge of long-context reasoning is fundamentally not a problem of context length, but of how information is searched, organized, and utilized during reasoning~\cite{CoA}.

Historically, efforts to enhance LLM performance on complex tasks have relied on a fixed-compute forward pass. Subsequent reasoning techniques, such as Chain-of-Thought (CoT)~\cite{CoT} and Tree-of-Thoughts (ToT)~\cite{ToT}, introduced the paradigm of \textit{compute scaling}~\cite{STTC} by increasing reasoning steps to deepen the model's internal reflection. Concurrently, Retrieval-Augmented Generation (RAG)~\cite{rag} represents a \textit{data scaling} approach, aiming to mitigate forgetting and hallucinations by incorporating external knowledge bases. While RAG mitigates catastrophic forgetting to some extent, na\"{i}ve retrieval paradigms remain fundamentally inadequate for resolving complex multi-hop reasoning challenges. The retrieved contexts frequently suffer from critical information omissions and semantic noise~\cite{asai2024selfrag}, leading to frequent failure in multi-hop long-context scenarios, as observed in benchmarks such as LoCoMo~\cite{LOCOMO} and LongMemEval~\cite{longmemeval}.

To specifically address long-context challenges, recent research has delved into AI Agent memory mechanisms~\cite{zhang2025survey}. While fundamental retrieval still relies on RAG-style pipelines, significant advancements have been made in the storage, segmentation, and construction of memory. 
Existing methods mainly include session-level memory~\cite{lu2023memochattuningllmsuse}, turn-level memory~\cite{mem0}, topic-aware memory~\cite{pan2025secom}, hierarchical summaries~\cite{MemoryBank,compassmem}, and graph-based retrieval structures such as RAPTOR~\cite{sarthi2024raptor} and HippoRAG~\cite{gutierrez2025hipporag}. Although these methods differ in how memory is organized and indexed, they share a common operational assumption that memory retrieval is treated as a single-step matching process~\cite{hu2026memoryageaiagents} between the current query and stored history. The agent must decide what to retrieve before knowing what information will become relevant during reasoning. Consequently, retrieval either becomes overly broad, introducing excessive irrelevant context that dilutes reasoning, or overly narrow, omitting crucial information and causing semantic fragmentation~\cite{rethinkingchunk}.

In essence, existing AI memory technologies encounter two fundamental bottlenecks in long-context multi-hop reasoning:
\textbf{(1) Search Modeling Dilemma in Long Contexts:} Extended texts are replete with dense entities and high-frequency noise. It is difficult to predict complex user intents a priori, causing static knowledge bases to suffer from severe semantic dilution during relational modeling.
\textbf{(2) Contextual Fragmentation:} Employing a single-granularity search strategy yields suboptimal results. Coarse-grained retrieval introduces excessive noise, whereas fine-grained retrieval strips away essential contextual background, leading to an imbalance between retrieval success rate and information density.
The root cause of these limitations is that current memory mechanisms treat retrieval as a static, single-step ``passive matching'' process. 

To this end, we go beyond the conventional view that memory merely assists reasoning, and instead propose that \textit{memory fundamentally defines the reasoning process}. This conceptualization drives a fundamental paradigm shift in how the agent interacts with its memory constraints. We introduce \textbf{MemCoT} (Memory-Driven Chain-of-Thought), a test-time scaling framework that reformulates long-context reasoning as an iterative memory–reasoning loop with evolving state.
This transforms memory access from static, single-shot passive matching~\cite{lewis2020rag} into a stateful, active search process~\cite{yao2023react}.
Within \textbf{MemCoT}, memory transcends the role of a passive data repository to function as a memory-driven reasoning engine, thereby rendering the reasoning process traceable and correctable. Our key contributions are summarized as follows:
\begin{itemize}
    \item
    We propose \textbf{MemCoT}, the first self-evolving architecture that couples test-time reasoning with dynamic memory evolution. By integrating a memory and reasoning loop, it effectively scales test-time compute, transforming static retrieval into an iterative search process that dynamically decomposes complex reasoning chains.
    
    \item 
    We design a multi-view long-term memory perception module with a trajectory- and context-aware short-term state. The perception module balances precise entity search with global contextual awareness, while the short-term state enables traceable reasoning that dynamically decomposes and prunes queries to condition memory access.

    \item
    MemCoT is a plug-and-play, training-free framework designed for complex long-context memory management, seamlessly empowering LLM agents in real-world scenarios. On the LoCoMo benchmark, it achieves SOTA results, enabling compact models like GPT-4o-mini and Qwen2.5-14B to reach $F_1$ scores of 58.03 and 57.06, respectively. Additionally, it achieves an 88.0 LLM-as-a-Judge score on the LongMemEval-s benchmark.
\end{itemize}

\section{Related Work}
\label{sec:related_work}

\subsection{Memory-Augmented Agents}

The integration of long-term memory mechanisms~\cite{zhang2025survey} marks a fundamental paradigm shift in the evolution of LLM agents~\cite{hu2026memoryageaiagents}. Early frameworks like MemGPT~\cite{memgpt} pioneered operating-system-inspired paging and segmentation to manage extended contexts. Building upon this, scalable architectures like Mem0~\cite{mem0} dynamically consolidate memory states to mitigate the severe limitations of fixed context windows. To elevate memory from a passive data repository to an active cognitive engine, recent architectures increasingly employ hierarchical and structured representations. Pioneering this structural shift, G-Memory~\cite{G-Memory} manages agent collaborations through a three-tier organizational graph hierarchy. To adaptively manage contextual noise within these agentic structures, approaches like MemGAS~\cite{MemGAS} propose dynamic multi-granularity routing. By transforming memory into a navigable logical map, architectures like Mnemis~\cite{Mnemis} empower agents to perform sophisticated cognitive traversals over long-horizon trajectories. Furthermore, addressing the computational overhead of complex memory constructions, frameworks like CoM~\cite{CoM} advocate for lightweight storage paired with sophisticated utilization, organizing memory fragments into coherent inference paths. Specialized structures are also emerging in complex multimodal domains, such as VisMem~\cite{VisMem}, which resolves visual processing bottlenecks via dual short-term and long-term memory systems.

\subsection{Evolutionary Memory}

Building upon persistent storage, the field is now decisively transitioning toward dynamic and self-evolving memory paradigms. Rather than maintaining static archives, recent research increasingly focuses on the continuous lifecycle and trajectory evolution of information. For instance, frameworks like CompassMem~\cite{compassmem} actively organize ongoing agent experiences into explicit event graphs, continually updating to map evolving causal and temporal sequences. Similarly, managing the decay and consolidation of these dynamic states is critical; MemoryBank~\cite{MemoryBank} selectively reinforces traits based on the Ebbinghaus Forgetting Curve to maintain consistency across prolonged interactions. Ultimately, frameworks are moving toward machine-native generative states: MemGen~\cite{memgen} introduces latent memory sequences that seamlessly interweave with reasoning, while MemSkill~\cite{memskill} and MemEvolve~\cite{MemEvolve} reframe memory operations as learnable, meta-adaptable skills that dynamically co-evolve with the agent's experiences and reasoning architecture.

\section{Motivation}
\label{sec:motivation}

A primary challenge in deploying LLM agents lies in managing memory during extended agent-environment interactions~\cite{zhang2025survey}.
Section \ref{sec:bayesian_modeling} formalizes the memory mechanism using Bayesian probability, highlighting the diminishing returns of one-step memory and the necessity of multi-step memory. Subsequently, Section \ref{sec:multi_view_retrieval} details our multi-view contextual retrieval design formulated to support this probabilistic model.

\begin{figure*}[t]
	\centering
	
	\begin{minipage}[t]{0.24\textwidth}
		\centering
		\vspace{0pt}
		\includegraphics[width=0.87\textwidth]{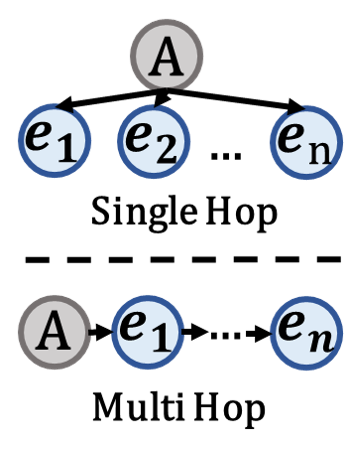}
		\subcaption{}
		\label{fig:mov1}
	\end{minipage}
	\hfill
	\begin{minipage}[t]{0.37\textwidth}
		\centering
		\vspace{0pt}
		\includegraphics[width=\textwidth]{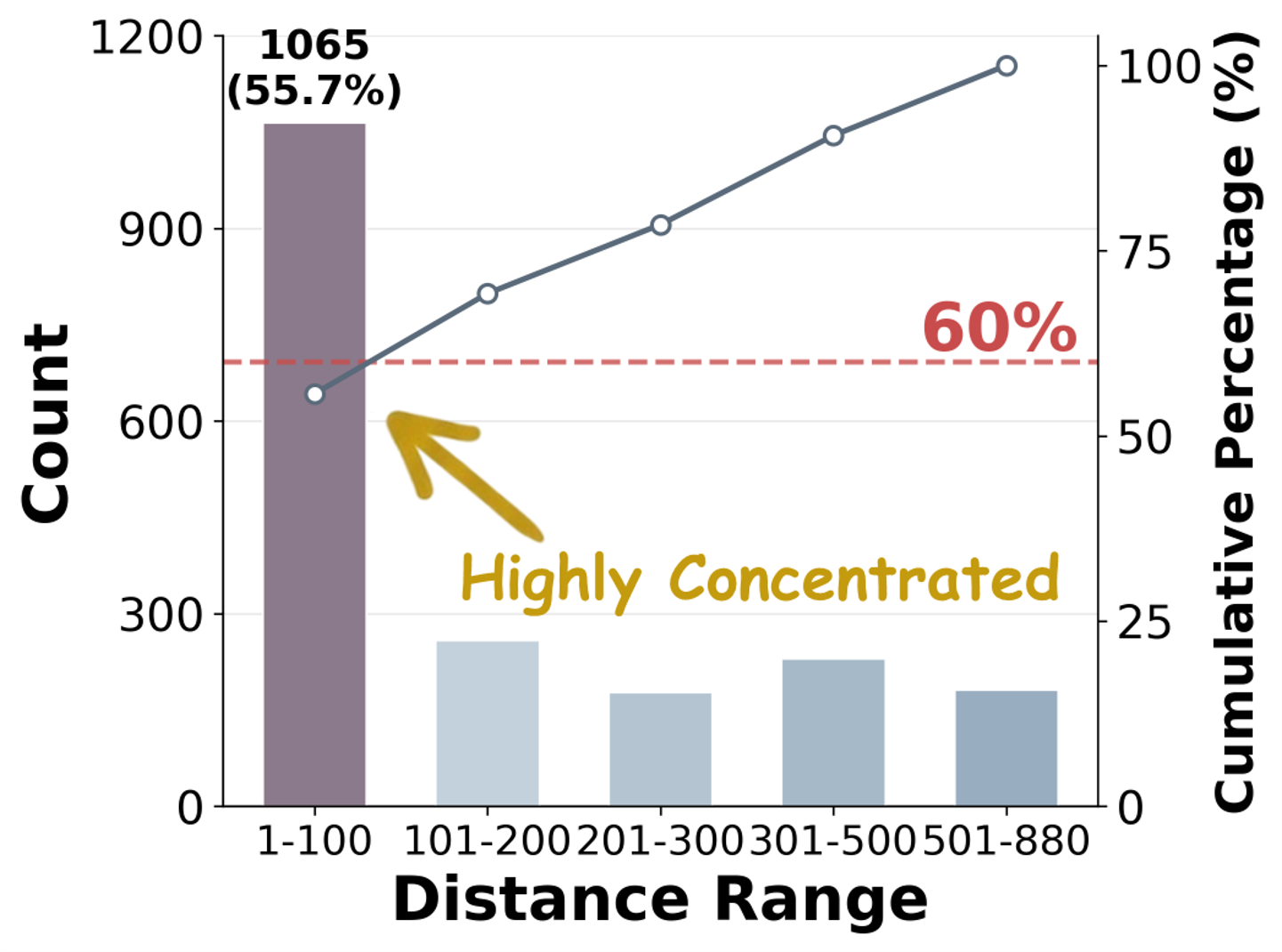}
		\subcaption{}
		\label{fig:mov2.1}
	\end{minipage}
	\hfill
	\begin{minipage}[t]{0.37\textwidth}
		\centering
		\vspace{0pt}
		\includegraphics[width=\textwidth]{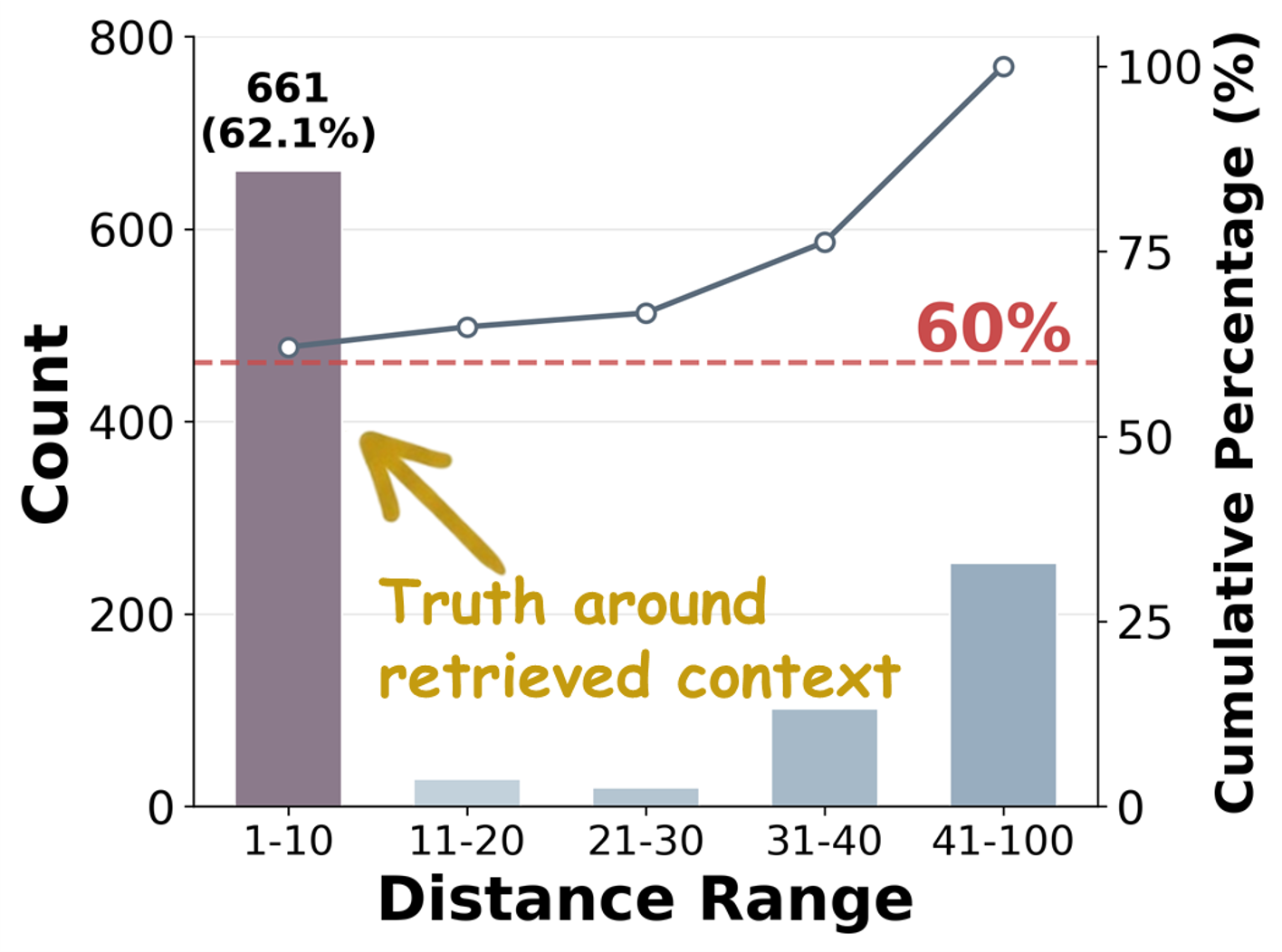}
		\subcaption{}
		\label{fig:mov2.2}
	\end{minipage}
    \vspace{-2mm}
	\caption{Analysis of reasoning and retrieval failures in long-context: (a) A relational diagram illustrating the dependencies between evidence and answers in complex single-hop and multi-hop reasoning scenarios. The red circle $A$ denotes the final answer, while the blue circles $e_i$ represent the supporting evidence. (b) A Pareto chart depicting the distribution of chunk distances (ranging from 1 to 880) between the falsely retrieved chunks and the ground-truth answer chunks evaluated on the LoCoMo dataset. (c) A magnified Pareto chart detailing the chunk distances within a narrowed range of 1 to 100 on the LoCoMo dataset.}
\end{figure*}
\subsection{Problem Formulation}
\label{sec:bayesian_modeling}

As illustrated in Figure \ref{fig:mov1}, both single-hop and multi-hop long-context queries require retrieving dispersed information. While existing one-step memory approaches handle simple tasks, they fail in complex scenarios: fixed top-$K$ bottlenecks limit single-hop retrieval, and high-frequency noise dilutes latent semantic links in multi-hop reasoning~\cite{MemAgent}. Fundamentally, modeling all potential reasoning paths within a single step is computationally intractable~\cite{liu-etal-2024-lost}. For a dataset $D$ and reasoning depth $n$, constructing a comprehensive knowledge graph $K'$ to map every potential query necessitates multiple Cartesian products: $K' = D \times D \times \dots \times D = D^n$. The state space of potential queries expands exponentially to $O(|D|^n)$. This prohibitive complexity inherently restricts the scalability of one-step methods in long-context scenarios~\cite{chen2025many}.

To address these bottlenecks, multi-step memory resolves complex intents through iterative searches~\cite{Evo-Memory}. We formalize this evolution as a Markov process. Given an initial query $Q$, required evidence $E$, and final answer $A$, optimizing long-context capability equates to maximizing the joint probability $q(A, E | Q)$. The following subsections decompose this objective for single-hop and multi-hop scenarios.
\\
\\
\noindent\textbf{Single-Hop Factorization.} 
For single-hop questions, particularly those requiring multiple pieces of evidence, we define the evidence set as $E = \{e_1, \dots, e_n\}$ and the corresponding answer components as $A = \{a_1, \dots, a_n\}$. Assuming conditional independence among the individual evidence-gathering events, we decompose the complex query $Q$ into $n$ atomic sub-queries, such that $Q = \{q_1, \dots, q_n\}$. The joint probability can thus be factorized sequentially:
\begin{equation}
\begin{aligned}
\tiny
    q(A, E | Q) &= q(a_1, e_1 | q_1) q(a_2, e_2 | q_2) \dots q(a_n, e_n | q_n) \\
    &= q(a_1 | e_1, q_1) q(e_1 | q_1) \dots q(a_n | e_n, q_n) q(e_n | q_n) \\
    &= \underbrace{\prod_{i=1}^{n} q(a_i | e_i, q_i)}_{\text{Generation Capacity}} \cdot \underbrace{\prod_{i=1}^{n} q(e_i | q_i).}_{\text{Memory Search Capacity}}
\end{aligned}
\label{eq:1}
\end{equation}

This derivation decouples the objective into response generation capacity of the responder agent, formulated as $q(A | E, Q) = f_{\theta}(E, Q)$, and memory search capacity, $q(e_i | q_i) = \psi(q_i, \mathcal{M})$, where $\psi(\cdot)$ denotes the memory retrieval function and $\mathcal{M}$ represents the input long-term memory. Because fixed top-$K$ limits restrict single-step retrieval, we decompose the $n$ required evidence pieces into $J$ sub-queries. Each targets $K=n/J$ pieces, formulating the search as a multi-step grouping process:
\begin{equation}
\begin{aligned}
\tiny
    \prod_{i=1}^{n} q(e_i | q_i) &= \prod_{j=1}^{J} \prod_{k=1}^{K} q(e_k | q_j, m_j) \\
    &= \prod_{j=1}^{J} q(e_{(j-1)\cdot K + 1}, \dots, e_{j \cdot K } | q_j, m_j) \\
    &= \prod_{j=1}^{J} q(E_j^{new} | q_j, m_j),
\end{aligned}
\label{eq:2}
\end{equation}
where $E_j^{new} = \{e_{(j-1)\cdot K + 1}, \dots, e_{j \cdot K} \}$.
Consequently, when addressing complex single-hop queries within extensive contexts, a one-step retrieval paradigm can at best resolve a mere fraction (e.g., $1/J$) of the information need, inevitably resulting in the truncation and loss of the remaining critical evidence. 
\\
\\
\noindent\textbf{Multi-Hop Exploration.}
For multi-hop reasoning, the required evidence set is modeled as $E = \{e_1\} \cup E_h$, where $e_1$ denotes the initial piece of evidence and $E_h = \{e_2, \dots, e_n\}$ represents the subsequent multi-hop evidence chain. Unlike the single-hop scenario, the events of acquiring each piece of evidence are no longer independent; given the initial query $Q$, there exists a strong sequential correlation among them. Based on the chain rule of probability, the joint probability must be factorized as follows:
\begin{equation}
\begin{aligned}
    q(A, E | Q) &= q(A, e_1, \dots, e_n| Q) \\
    &= \underbrace{q(A |e_{n} , \dots, e_1, Q)}_{\text{Generation Capacity}} \underbrace{q(e_1, \dots, e_n| Q).}_{\text{Memory Search Capacity}}
\end{aligned}
\label{eq:3}
\end{equation}

In this sequential modeling, the final term $q(A | e_n, \dots, e_1, Q) = f_{\theta}(E, Q)$ still represents the responder agent's generation capacity based on all accumulated evidence. The query $Q$ can still be decomposed into sub-problems, but their resolutions are strictly conditioned on prior discoveries. Mathematically, the multi-hop memory search capacity is also modeled as a multi-step grouping process:
\begin{equation}
    \begin{aligned}
        q(e_1, \dots, e_n| Q)=q(e_1 | Q) q(e_2 | e_1, Q) \dots q(e_{n} | e_{n-1}, \dots,  e_1, Q).
    \end{aligned}
    \label{eq:4}
\end{equation}
Consequently, for complex multi-hop reasoning tasks, a one-step search is prone to extracting superficial evidence, culminating in vague and ungrounded final responses.

In summary, static one-step retrieval is highly impractical; mastering long-context tasks requires a dynamic, multi-step evolving memory architecture.
\subsection{Perceptual Scale Mismatch for Retrieval}
\label{sec:multi_view_retrieval}

RAG systems are notoriously sensitive to chunking strategies~\cite{guo-etal-2025-lightrag, LOCOMO}. Oversized chunks precipitate the omission of critical entities and relations, leading to semantic ambiguity during retrieval. Conversely, undersized chunks restrict the retrieval's informational horizon, causing contextual fragmentation~\cite{rethinkingchunk}.

In extended-context environments, target information is frequently dispersed asymmetrically across the dataset~\cite{rethinkingchunk}. Therefore, a retrieval methodology that guarantees both precise semantic targeting and a broad informational horizon is highly valuable~\cite{rethinkingchunk}. To empirically investigate this necessity, we analyzed the retrieval failure cases of LightRAG~\cite{guo-etal-2025-lightrag} on the long context conversation benchmark LoCoMo. Specifically, we quantified the positional distance between the falsely retrieved chunks and the ground-truth chunks containing the correct answers. As illustrated in Figure~\ref{fig:mov2.1}, when employing a smaller chunk setting (approximately 200 tokens per chunk), roughly 60\% of the erroneous retrievals are located within a 100-chunk radius of the actual answer. Furthermore, as depicted in Figure~\ref{fig:mov2.2}, approximately 60\% of these proximal results are concentrated within a mere 10-chunk distance. 
This empirical analysis demonstrates that requisite answers are frequently distributed within the immediate periphery of the initially retrieved information~\cite{rethinkingchunk,deepseekengram}. Motivated by these findings, we propose a multi-view design paradigm characterized by fine-grained small-vision retrieval, combined with coarse-grained wide-vision observation to optimize memory reading operations.

\begin{figure*}[htbp]
    \centering
    \includegraphics[width=1\textwidth]{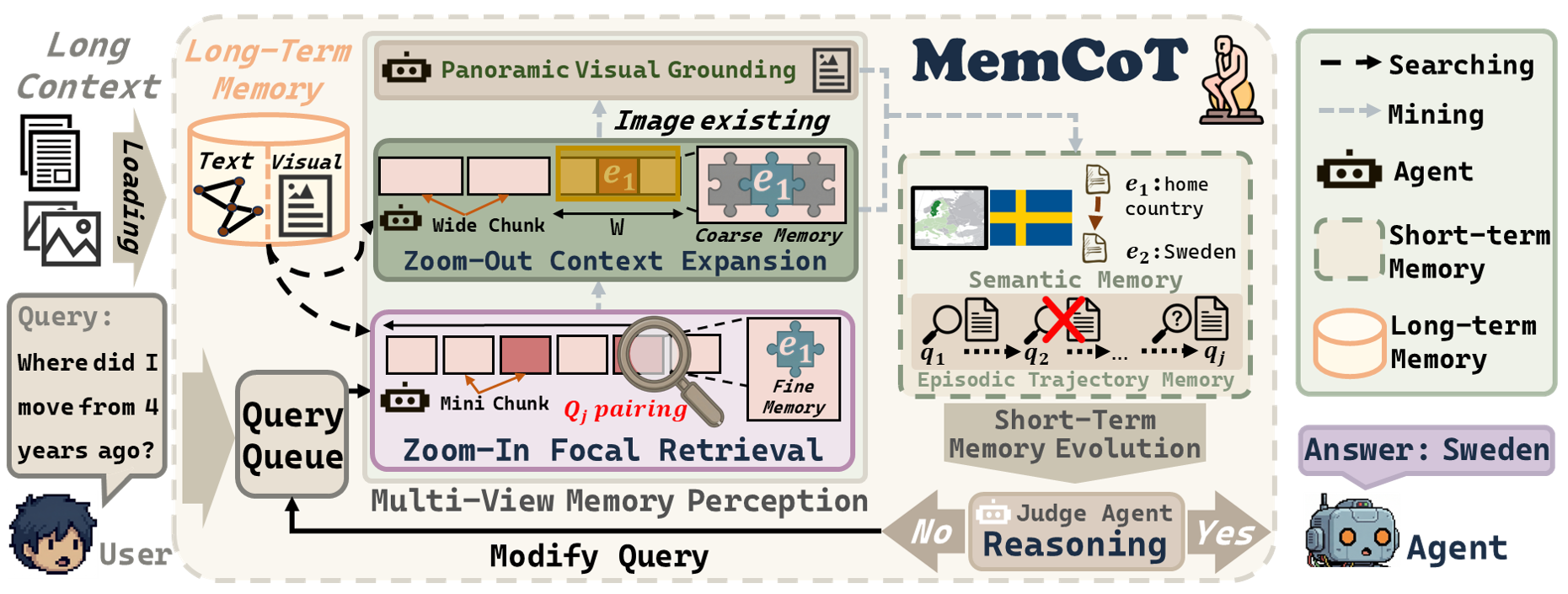}
    \caption{The overall architecture of the MemCoT framework. It illustrates the scaling test-time compute memory loop, driven by a multi-view memory perception module for long-term memory extraction and a short-term memory evolution module for iterative query $q_j$ updates in the $j-$th iteration. On one hand, the retrieved memory $m_{j}$ provides factual anchors that empower the Judge Agent to construct a more rigorous and grounded CoT for reasoning. On the other hand, the reasoning process explicitly identifies the deficiencies in the retrieved memory $m_{j}$ relative to the current query $q_j$ and refines the $q_j$ into the next query $q_{j+1}$ to optimize the next retrieved memory $m_{j+1}$. This bidirectional synergy loop establishes the foundation of our \textit{Memory-driven Chain of Thought} paradigm.}
    \label{fig:main}
\end{figure*}
\section{Methodology}
\label{sec:method}

In this section, we present the framework of MemCoT. In Section \ref{sec:framework}, we introduce the scaling test-time compute memory framework in MemCoT, detailing the self-evolution of reasoning and memory. Section \ref{sec:long_term} describes the multi-view long-term memory perception within MemCoT. Finally, Section \ref{sec:short_term} elaborates on the task-conditioned dual short-term memory system.

\subsection{Overview of MemCoT Framework}
\label{sec:framework}
MemCoT facilitates test-time memory acquisition and state evolution through a recurrent memory-reasoning loop. As illustrated in Figure \ref{fig:main}, each iteration $j$ begins with the perception module $\psi_{perceive}(q_j, \mathcal{M})$, which performs deep extraction over the long-term memory $\mathcal{M}$. Subsequently, the evolution module $f_\theta^{\text{evolve}}$ distills high-value semantic and episodic trajectory information, updating the dynamic short-term memory state $m_j$. The judge process $f_\theta^{\text{judge}}$ evaluates the sufficiency of $m_j$: if adequate, the responder agent generates the final answer; otherwise, $f_\theta^{\text{judge}}$ identifies ``knowledge blind spots'' to formulate the next sub-query $q_{j+1}$, triggering a new cycle. This interplay effectively decomposes multi-hop reasoning into tractable sub-problems. According to Sec~\ref{sec:bayesian_modeling}, we observe that both single-hop and multi-hop retrieval paradigms can be uniformly modeled through an iterative process. Both paradigms are encapsulated into a single, unified equation:
\begin{equation}
    q(A, E | Q) = f_{\theta}(E, Q) \prod_{j=1}^{J} f_\theta^{\text{memcot}}(q_j, m_j, Q),
    \label{eq:5}
\end{equation}
where $f_\theta^{\text{memcot}}$ extracts contextually useful information by evaluating whether the current memory state $m_j$ adequately addresses the sub-query $q_j$ and the overarching user query $Q$. For each iteration $j = 1, \dots, J$, the state transitions rely on the following equations:
\begin{align}
        m_j &= f_\theta^{\text{evolve}}(m_{j-1}, \psi_{perceive}(q_j, \mathcal{M}), q_j), \label{eq:6} \\
        q_{j+1} &= f_\theta^{\text{judge}}(m_j, q_j),
        \label{eq:7}
\end{align}
where $q_1=Q$ and $m_{0}=\emptyset$.
This mathematically unified framework demonstrates the theoretical 
versatility of the proposed memory evolution mechanism. The detailed mathematical derivation and proof for this unification are provided in Appendix A.

\subsection{Multi-View Long-Term Memory Perception}
\label{sec:long_term}

This module consists of a multi-modal long-term memory storage unit, denoted as $\mathcal{M}$, and a multi-view memory perception function $\psi_{perceive}$. To effectively balance retrieval precision and contextual completeness, we design $\psi_{perceive}$ as a hierarchical aggregation of zoom-in focal retrieval $\psi_{\text{z-i}}$, zoom-out context expansion $\psi_{\text{z-o}}$, and panoramic visual grounding $\psi_{\text{vis}}$. Formally, for a given sub-query $q_j$ at the $j$-th evolutionary iteration, the multi-view perception process is mathematically formulated as:
\begin{equation}
\begin{aligned}
    m_{\text{z-i}} &= \psi_{\text{z-i}}(q_j, \mathcal{M},K), \\
    m_{\text{z-o}} &= \psi_{\text{z-o}}(q_j, \mathcal{M}, m_{\text{z-i}}, W), \\
    \psi_{perceive}(q_j, \mathcal{M}) &= \underbrace{m_{\text{z-i}} \cup m_{\text{z-o}}}_{\text{ Text Type}} \cup \underbrace{\psi_{\text{vis}}(q_j, \mathcal{M}, m_{\text{z-i}}),}_{\text{ Visual Type}}
\end{aligned}
\end{equation}
where $\cup$ denotes the aggregation of retrieved multi-view memory contents.

For text-type memory, we adhere to the core design principle of ``fine-grained search, multi-view mining.'' Built upon a graph-structured storage framework (e.g., LightRAG~\cite{guo-etal-2025-lightrag}), the memory entries in $\mathcal{M}$ are partitioned into minimal textual chunks to maximize the granularity of entities and relations. During the initial retrieval phase, the zoom-in focal retrieval function $\psi_{\text{z-i}}$ extracts the highly precise and albeit contextually narrow chunks from top-$K$ memory chunks to form $m_{\text{z-i}}$.  To mitigate the inherent narrowness of this receptive field, the zoom-out context expansion function, $\psi_{\text{z-o}}$, is subsequently triggered. By leveraging the positional indices of $m_{\text{z-i}}$, it retrieves an adjacent window of $W$ chunks, yielding a wide viewing memory $m_{\text{z-o}}$ representation that successfully restores the surrounding contextual semantics.

When the long-term memory encompasses visual-type memory, we employ a unified multi-modal structural memory approach. Recognizing that the spatial proximity between images and their surrounding text encapsulates subtle visual topologies~\cite{deepseekocr}, we leverage Optical Character Recognition (OCR) methodologies to project images and adjacent text into a unified visual representation. If the wide viewing memory $m_{\text{z-o}}$ retrieval detects an explicit image cue, the system seamlessly triggers the panoramic visual grounding module, $\psi_{\text{vis}}$. This function retrieves the corresponding multi-modal memory to extract macro-structural information, thereby comprehensively optimizing the final response quality. To minimize computational latency and token overhead, the image resolutions are dynamically compressed during this visual extraction phase.

Ultimately, according to Eqs.~(\ref{eq:6}) and (\ref{eq:7}), the aggregated high-value information extracted across these varied views constitutes the semantic memory. This refined semantic representation is then explicitly written into the dynamic short-term memory state $m_j$ for subsequent reasoning and query updating.

\subsection{Short-Term Memory Evolution}
\label{sec:short_term}

A dynamic short-term memory is the critical nexus for state evolution, directly governing reasoning efficacy. To transform retrieval from passive matching to active perception, we introduce a task-conditioned evolving memory $f_\theta^{\text{evolve}}$, which integrates semantic memory and episodic trajectory memory within the dynamic short-term memory $m_j$.

As shown in the right part of Figure \ref{fig:main}, after $f_\theta^{\text{evolve}}$ updates $m_{j-1}$ to $m_j$, the new semantic memory stored in $m_j$ contains the factual semantic memory returned by the long-term perception module $\psi_{perceive}$. Meanwhile, the new episodic trajectory memory records the state transitions and failed query trajectories across iterations. During each cycle, the judge agent evaluates the dynamic short-term memory $m_j$ during the judge process $f_\theta^{\text{judge}}$. If the $m_j$ contains sufficient evidence, the evolution stops and returns $m_j$ to the responder agent. If not, according to Eq. (\ref{eq:7}), the modifying query agent analyzes the failed trajectories to understand the reasoning gap and dynamically updates the query $q_j$ as $q_{j+1}$.

The query state update consists of two primary actions:
\begin{itemize}
    \item \textbf{Decomposition:} For complex intents, the modifying query agent dissects the query into atomic sub-queries (e.g., decomposing ``What did John do in England and Japan 4 years ago?'' into separate queries for ``John do in England 4 years ago.'' and ``John do in Japan 4 years ago.'').
    \item \textbf{Pruning:} If the semantic memory has already resolved a portion of the query, the modifying query agent deletes the resolved intent to focus strictly on the missing evidence.
\end{itemize}
By continuously referencing the failed trajectories within the episodic trajectory memory, the modifying query agent prevents redundant searches. This targeted state evolution ensures that subsequent queries are tightly aligned with the reasoning deficits, maximizing the efficacy of long-term memory retrieval.

\begin{table*}[htbp]
\centering
\setlength{\tabcolsep}{4mm}
\caption{
Performance  ($F_1$ score) comparison on the LoCoMo benchmark, covering single-hop, multi-hop, open-domain, and temporal settings. Best results are highlighted in \textbf{bold}, and second-best results are \underline{underlined}.
}
\begin{tabular}{ll ccccc}
\toprule
\textbf{Model} & \textbf{Method}
& \textbf{Single-hop}& \textbf{Multi-hop}& \textbf{Open-domain}& \textbf{Temporal}& \textbf{Overall}\\
\midrule

\multirow{9}{*}{\rotatebox{90}{\textbf{GPT-4o-mini~\cite{4omini}}}}
& Full Context& 45.89& 25.01& 15.49& 23.83& 35.57\\
 & RAG~\cite{rag} & 52.19 & 32.17 & 23.21 & 30.77 &42.25 \\
& Mem0~\cite{mem0}& 47.65 & 38.72 & \underline{28.64} & 48.93 & 45.10 \\
& MemoryOS~\cite{li2025memosmemoryosai} & 48.62 & 35.27 & 20.02 & 41.15 & 42.84 \\
& HippoRAG~\cite{gutierrez2025hipporag} & 54.84 & 33.59 & 28.59 & 48.17 & 47.92 \\
& A-Mem~\cite{xu2025amem} & 44.65 & 27.02 & 12.14 & 45.85 & 39.65 \\
& CAM~\cite{li2025cam} & 50.58 & 33.55 & 18.23 & 44.14 & 44.10 \\
& CompassMem~\cite{compassmem}& \underline{57.36}& \underline{38.84}& 26.61& \underline{57.96}& \underline{52.18}\\
& \cellcolor[RGB]{235,245,250}MemCoT & \cellcolor[RGB]{235,245,250}\textbf{64.81} & \cellcolor[RGB]{235,245,250}\textbf{40.43} & \cellcolor[RGB]{235,245,250}\textbf{42.67} & \cellcolor[RGB]{235,245,250}\textbf{60.31} & \cellcolor[RGB]{235,245,250}\textbf{58.03} \\
\midrule
\multirow{9}{*}{\rotatebox{90}{\textbf{Qwen2.5-14B~\cite{qwen2025qwen25technicalreport}}}}& Full Context& 55.35& 32.59& 13.50& 27.36& 42.74\\
 & RAG~\cite{rag}& 49.79 & 28.11 & 20.42 & 24.73 &38.77 \\
& Mem0~\cite{mem0} & 42.58 & 31.73 & 15.03 & 28.96 & 36.04 \\
& MemoryOS~\cite{li2025memosmemoryosai} & 46.33 & 38.19 & 20.27 & 32.24 & 40.28 \\
& HippoRAG~\cite{gutierrez2025hipporag} & 42.45 & 27.57 & 19.74 & 30.66 & 35.85 \\
& A-Mem~\cite{xu2025amem} & 33.75 & 22.09 & 13.49 & 27.19 & 28.98 \\
& CAM~\cite{li2025cam} & 50.39 & 34.50 & 23.86 & 44.70 & 44.64 \\
&CompassMem~\cite{compassmem}&\underline{61.02}& \underline{42.32}& \underline{25.88}& \underline{47.18} & \underline{52.52} \\
& \cellcolor[RGB]{235,245,250}MemCoT & \cellcolor[RGB]{235,245,250}\textbf{63.09} & \cellcolor[RGB]{235,245,250}\textbf{45.10} & \cellcolor[RGB]{235,245,250}\textbf{38.04} & \cellcolor[RGB]{235,245,250}\textbf{57.49} & \cellcolor[RGB]{235,245,250}\textbf{57.06} \\
\bottomrule
\end{tabular}
\label{tab:main}
\end{table*}
\section{Experiment}
\label{sec:experiment}

In this section, we comprehensively evaluate the MemCoT framework. We first introduce the experimental settings, followed by the main results on the long-context benchmark LoCoMo and LongMemEval~\cite{LOCOMO,longmemeval}. Finally, we conduct ablation studies and hyperparameter analyses to validate the efficacy of our core components.

\subsection{Experimental Settings}
\label{ssec:exp_settings}

\textbf{Datasets.}
We evaluate our approach on two primary datasets designed to assess long-range memory and reasoning capabilities in agent systems. First, we utilize the LoCoMo benchmark~\cite{LOCOMO}, which comprises 841 single-hop, 282 multi-hop, 96 open-domain, and 321 temporal reasoning challenges. We report the $F_1$ score as the primary evaluation metric across all sub-tasks. Second, we incorporate the LongMemEval-S benchmark\cite{longmemeval} to further validate comprehensive memory performance across distinct evaluation dimensions. For the LongMemEval-S benchmark~\cite{longmemeval}, we report the overall performance across 70 sessions, alongside specific granular metrics comprising 133 Single-session-user (SSU), 30 Multi-session (MS), 30 Single-session-preference (SSP), 133 Temporal-reasoning (TR), 78 Knowledge-update (KU), and 56 Single-session-assistant (SSA) instances.

\textbf{Implementation Details.}
To ensure a robust evaluation, we utilize both closed-source and open-source models as our foundational responders: GPT-4o-mini~\cite{4omini} and Qwen2.5-14B~\cite{qwen2025qwen25technicalreport}. For LongMemEval-S~\cite{longmemeval}, $W$ is increased to 15 to capture the fragmented dependencies inherent in its multi-session structures. Our retrieval module employs LightRAG with the recommended Top-$K=10$ for focal retrieval. The maximum evolutionary iterations $J$ is uniformly set to 8 across all tasks to facilitate deep reasoning while preventing indefinite search cycles.

\begin{table*}[htbp]
\centering
\setlength{\tabcolsep}{4mm}
\caption{Performance ($F_1$ score) comparison on the LoCoMo benchmark using the tiny model Qwen2.5-7B model, highlighting MemCoT's consistent superiority even in smaller-scale architectures.}
\begin{tabular}{ll ccccc}
\toprule
 \textbf{Model}&  \textbf{Method}& \textbf{Single-Hop} & \textbf{Multi-Hop} &  \textbf{Open-domain} &\textbf{Temporal} & \textbf{Overall} \\
\midrule
\multirow{9}{*}{\rotatebox{90}{\textbf{Qwen2.5-7B~\cite{qwen2025qwen25technicalreport}}}}& Full Context& 10.61 & 3.39 &  8.24 &3.72 & 7.38 \\
& RAG~\cite{rag}& 21.72 & 8.05 &  14.51 &4.89 & 13.88 \\
& A-Mem~\cite{xu2025amem}& 12.73 & 8.67 &  13.86 &6.28 & 12.1 \\
& EpMem~\cite{EpMem}& 17.64 & 6.35 &  2.66 &9.96 & 6.63 \\
& Mem0~\cite{mem0}& 24.96 & 20.31&  32.74 &\underline{33.16}& 28.75 \\
& MemoryOS~\cite{li2025memosmemoryosai}& \underline{29.55}& 21.03 &  \textbf{40.85} &26.26& 33.74 \\
& MemVerse~\cite{memverse} & 28.12 & \underline{24.23}&  \underline{40.33} &25.05 & \underline{33.79}\\
 & \cellcolor[RGB]{235,245,250}MemCoT & \cellcolor[RGB]{235,245,250}\textbf{63.99}& \cellcolor[RGB]{235,245,250}\textbf{36.34}&  \cellcolor[RGB]{235,245,250}33.08&\cellcolor[RGB]{235,245,250}\textbf{40.06}&\cellcolor[RGB]{235,245,250}\textbf{52.01}\\
\bottomrule
\end{tabular}
\label{tab:small_model}
\end{table*}
\begin{table*}[htbp]
\centering
\setlength{\tabcolsep}{5mm}
\caption{
Performance  (LLM-as-a-Judge score) comparison on the LongMemEval-S benchmark. Best results are highlighted in \textbf{bold}, and second-best results are \underline{underlined}.
}
\begin{tabular}{ll ccccccc}
\toprule
\textbf{Model} & \textbf{Method}
& \textbf{SSU} & \textbf{MS} & \textbf{SSP} & \textbf{TR} & \textbf{KU} & \textbf{SSA} & \textbf{Overall}\\
\midrule
\multirow{7}{*}{\rotatebox{90}{\textbf{GPT-4o-mini~\cite{4omini}}}}
& Full Context & 78.6 & 38.3 & 6.7 & 42.1 & 78.2 & 89.3 & 55.0 \\
& RAG~\cite{rag} & 88.6 & 47.4 & 70.0& 63.2 & 70.5 & 91.1 & 67.2 \\
& Mem0~\cite{mem0} & 91.4 & 66.2 & 34.0 & 63.9 & 74.4 & \underline{96.4} & 71.1 \\
& Zep~\cite{zep} & 92.9& 47.4 & 53.3 & 54.1 & 74.4 & 75.0 & 63.2 \\
& Nemori~\cite{Nemori} & 88.6 & 51.1 & 46.7 & 61.7 & 61.5 & 83.9 & 64.2 \\
& EMem-G~\cite{EMem-G} & 87.0 & 73.6& 32.2 & 74.8& \textbf{94.4} & 87.5 & 77.9\\
& Mnemis~\cite{Mnemis} & \underline{97.1}& \underline{76.7}& \textbf{90.0} & \underline{83.5}& \underline{92.3} & \textbf{100.0} & \underline{87.2}\\
& \cellcolor[RGB]{235,245,250}MemCoT & \cellcolor[RGB]{235,245,250}\textbf{98.5}& \cellcolor[RGB]{235,245,250}\textbf{79.6}& \cellcolor[RGB]{235,245,250}\underline{70.0}& \cellcolor[RGB]{235,245,250}\textbf{89.4}& \cellcolor[RGB]{235,245,250}88.4& \cellcolor[RGB]{235,245,250}\textbf{100.0}& \cellcolor[RGB]{235,245,250}\textbf{88.0}\\ 
\bottomrule
\end{tabular}
\label{tab:longmemeval}
\end{table*}

\begin{figure*}[htbp]
    \centering
    \includegraphics[width=0.9\textwidth]{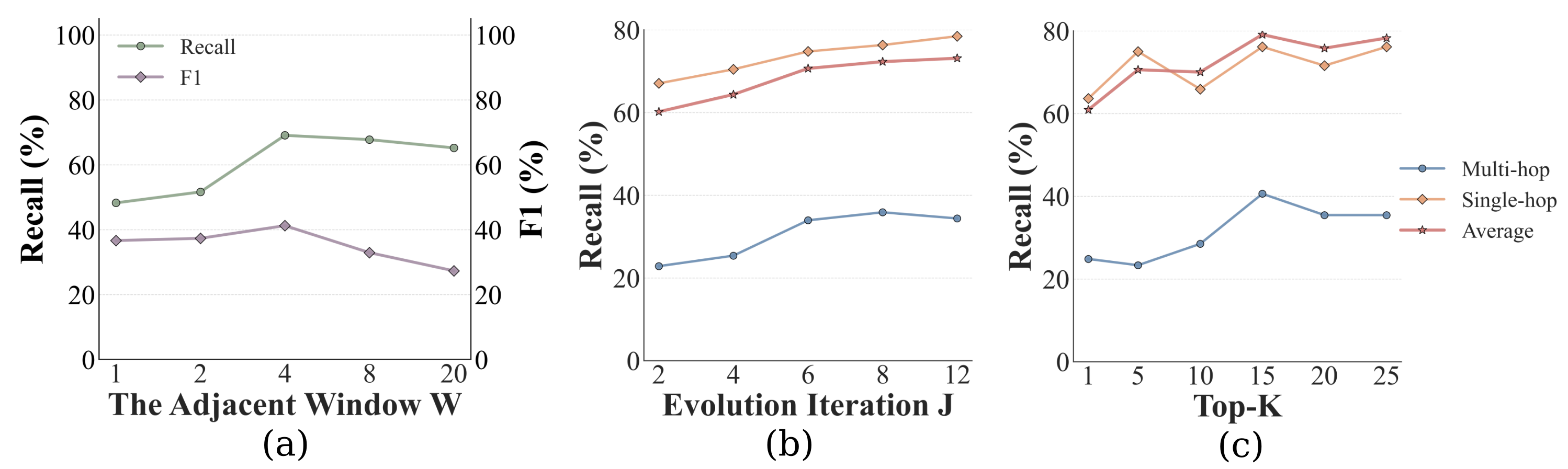}
    
    \caption{Effects of different parameters: (a) Impact of the adjacent window size ($W$) on memory retrieval Recall (\%) and reasoning $F_1$ score (\%) using the Qwen2.5-7B model. (b) Impact of the maximum evolution iterations ($J$) on memory retrieval Recall (\%) using the Qwen2.5-7B model. (c) Impact of Fine-grained TopK ($K$) on memory retrieval Recall (\%) using the Qwen2.5-7B model.}
    \label{fig:JK}
\end{figure*}

\subsection{Main Results}
\label{ssec:main_results}

The main comparative results on the LoCoMo benchmark are presented in Table \ref{tab:main} and Table \ref{tab:small_model}. Across all tested architectures, including GPT-4o-mini, Qwen2.5-14B, and Qwen2.5-7B, our proposed MemCoT framework consistently establishes new SOTA results.

Specifically, on the GPT-4o-mini model, MemCoT achieves a commanding overall $F_1$ score of 58.03, significantly outperforming the strongest graph-based baseline, CompassMem, which scores 52.18. This substantial improvement is particularly pronounced in complex reasoning scenarios such as Temporal (60.31) and Open-domain (42.67) questions, demonstrating the framework's superior capability in dynamic state evolution and active perception. Similarly, when equipped with the Qwen2.5-14B model, MemCoT attains an impressive overall $F_1$ score of 57.06, compared to CompassMem's 52.52 and CAM's 44.64. Moreover, it is noteworthy that while Table~\ref{tab:main} highlights a substantial performance gain in open-domain reasoning, this surge is not proportionally reflected in the aggregate overall score. We attribute this phenomenon to the class imbalance within the LoCoMo benchmark, as the open-domain category comprises only 96 instances, thereby exerting a limited influence on the final weighted average.

To validate robustness, we evaluated MemCoT on the compact Qwen2.5-7B. As shown in Table \ref{tab:small_model}, MemCoT achieves a 52.01 $F_1$ score, representing a 57.3\% improvement over the strongest baseline MemVerse (33.79). Notably, despite limited parameters, MemCoT maintains a strong score for the multi-hop question (36.34), single-hop question (63.99), and temporal question (40.06). This success stems from our memory evolution mechanism: by using GPT-4o-mini for high-quality memory construction, MemCoT enables a test-time knowledge distillation effect, allowing smaller models to internalize complex structural insights.

Furthermore, the evaluation on the LongMemEval-S benchmark corroborates the comprehensive superiority of our approach, as detailed in Table \ref{tab:longmemeval}. Using the GPT-4o-mini model, MemCoT achieves the highest overall score of 88.0, surpassing strong, specialized long-term memory baselines such as Mnemis (87.2) and EMem-G (77.9). The framework demonstrates exceptional proficiency in extracting and tracking conversational details, yielding SOTA results in Single-session-user (SSU) with 98.5, Multi-session (MS) with 79.6, and Temporal-reasoning (TR) with 89.4. Additionally, MemCoT perfectly executes Single-session-assistant (SSA) tracking with a flawless score of 100.0. Despite competitive niche performances by EMem-G and Mnemis in KU and SSP, MemCoT’s superior overall results underscore its versatile capacity for multi-dimensional memory integration.

\begin{table}[t]
\centering
\setlength{\tabcolsep}{1mm}
\caption{Ablation study results on the LoCoMo benchmark using Qwen2.5-14B ($F_1$ scores). $\psi_{\text{z-o}}$ and $\psi_{\text{vis}}$ denote zoom-out and visual modules, respectively.}
\begin{tabular}{lccccc}
\toprule
Method& Multi-hop & Temporal & Open & Single-hop & Overall \\
\midrule
MemCoT &40.43 & 60.31& 42.67& 64.81& 58.03\\
\midrule
w/o $\psi_{\text{z-o}}$ & 35.48
& 55.13
& 38.58
& 61.35
& 53.89\\
w/o $\psi_{\text{vis}}$ & 39.21
& 59.73
& 39.58
& 63.82
& 56.19 \\
w/o $\psi_{\text{z-o+vis}}$ & 34.78
& 50.88
& 41.49
& 59.62
& 53.71 \\
\bottomrule
\end{tabular}
\label{tab:ablation}
\end{table}

\subsection{Ablation Study}
\label{ssec:ablation}

To systematically evaluate the contributions of the core mechanisms within MemCoT, we design several ablation variants.

\textbf{Multi-view Long-Term Memory Perception.}
Our perception module integrates coarse-grained and global extraction to balance precision and contextual integrity. We investigate the impact of these components by testing the following variants:
\begin{itemize}
    \item \textbf{w/o $\psi_{\text{z-o}}$}: We remove the zoom-out context expansion module. Relying solely on fine-grained chunks restricts the informational horizon and is expected to exacerbate contextual fragmentation, leading to performance degradation on complex multi-hop queries.
    \item \textbf{w/o $\psi_{\text{vis}}$}: We disable the panoramic visual grounding module. This variant isolates the text-type memory from multi-modal structural cues, demonstrating the necessity of macro-structural visual integration for comprehensive response quality.
\end{itemize}
As shown in Table~\ref{tab:ablation}, removing $\psi_{\text{z-o}}$ leads to a substantial drop in $F_1$ scores. In contrast, the decline caused by ablating the $\psi_{\text{vis}}$ is comparatively mild on the LoCoMo benchmark.

\subsection{Further Analysis}

We further conduct sensitivity analyses on key hyperparameters that govern the dynamic short-term memory evolution and multi-view retrieval processes.

\textbf{Impact of Adjacent Window Size ($W$).} 
To further elucidate the role of contextual breadth in our framework, we investigate the sensitivity of the adjacent window size $W$ within the zoom-out context expansion module $\psi_{\text{z-o}}$. As illustrated in Figure~\ref{fig:JK}(a), when $W$ scales from 1 to 4, both the retrieval Recall and the final $F_1$ score exhibit a consistent upward trajectory. This trend confirms that $\psi_{\text{z-o}}$ successfully enlarges the model's informational horizon, effectively mitigating the contextual fragmentation typically associated with fine-grained retrieval by restoring surrounding semantic dependencies. However, as $W$ exceeds 8, the Recall rate plateaus while the $F_1$ score begins a pronounced decline. We attribute this performance degradation to the fact that an excessively large $W$ introduces a surplus of retrieved memory entries, which over-extends the input context and incorporates significant semantic noise. This influx of irrelevant information dilutes critical signals and triggers reasoning instability, ultimately compromising the quality and precision of the generated responses.

\textbf{Impact of Maximum Evolutionary Steps ($J$).}
The hyperparameter $J$ constrains the maximum allowable number of iterative search cycles to prevent indefinite trajectories. In Figure~\ref{fig:JK}(b), the Judge Agent autonomously determines when sufficiently accurate information has been gathered in the short memory; consequently, varying $J$ within the range of 2 to 6 leads to substantial fluctuations in recall rate, while values beyond 6 result in only marginal variations. 

\textbf{Impact of Fine-grained TopK ($K$).}
The variable $K$ serves as the lower bound of evidence acquisition per retrieval step during the initial zoom-in focal retrieval module $\psi_{\text{z-i}}$. As illustrated in Figure~\ref{fig:JK}(c), setting $K$ in the range of 10–15 yields a notable performance gain, whereas values exceeding 15 lead to a decline. We attribute this to the increased number of retrieved memory entries expanding the context of $\psi_{\text{z-i}}$ excessively, thereby inducing catastrophic forgetting.

\section{Conclusion}
In this paper, we introduce MemCoT, a test-time memory scaling framework that redefines memory as a memory-driven engine for long-context LLM reasoning. By integrating a recurrent memory-agent loop, MemCoT transforms static retrieval into an active, stateful search process. Our architecture synergizes multi-view long-term perception with a dynamic short-term memory system to systematically decompose complex queries and prune redundant trajectories. Extensive evaluations demonstrate that MemCoT achieves new SOTA performance on the LoCoMo and LongMemEval-S benchmarks. Ultimately, MemCoT provides a mathematically unified and scalable solution for overcoming both single-hop and multi-hop reasoning bottlenecks in extended interactions. Looking ahead, \textbf{MemCoT} facilitates a fundamental paradigm shift from passive retrieval to active, stateful cognitive reasoning, empowering AI agents to maintain rigorous coherence throughout extended interactions. This architecture establishes a mathematically unified and scalable foundation for the next generation of self-evolving, long-context intelligent systems.

\bibliographystyle{ACM-Reference-Format}
\bibliography{sample-base}

%
\clearpage
\appendix
\section{Unified Objective in MemCoT} 
\label{app:proof-eq5}

This appendix formalizes how the unified objective in Eq.~\eqref{eq:5} follows from the single-hop decomposition (Eqs.~\eqref{eq:1}--\eqref{eq:2}) and the multi-hop chain factorization (Eqs.~\eqref{eq:3}--\eqref{eq:4}), once the MemCoT loop in Eqs.~\eqref{eq:6}--\eqref{eq:7} is interpreted as a sequential memory search process.

For a single-hop query, to solve the modeled grouping, our MemCoT conducts $J$ rounds of search for the evidence (from $j=1$ to $J$). During each evolutionary round $j$, it forms a short-term memory $m_j$ and a corresponding query $q_j$. We expect that each short-term memory evolution incorporates at least $K$ new pieces of evidence (i.e., $|E_j^{\text{new}}| = K$). In a single round of search, the single-hop formation is defined as:
\begin{equation}
    q(E_j^{\text{new}} | q_j, m_j) = f_\theta^{\text{memcot}}(q_j, m_j, Q)
\end{equation}

For a multi-hop query, to resolve the complex conditional dependency, the probability of retrieving the $j$-th hop evidence is flawlessly modeled by the state transition of our short-term memory. The initial step of this process is defined as $q(e_1 | Q) = f_\theta^{\text{memcot}}(q_1, \\m_0, Q)$ with $q_0 = Q$ and $m_{0} = \emptyset$. Subsequently, each conditional step is defined as:
\begin{equation}
    q(e_j | e_{j-1}, \dots, e_1, Q) = f_\theta^{\text{memcot}}(q_j, m_j, Q)
\end{equation}

By substituting the iterative state transitions back into our prior formulations, we demonstrate that both problem types converge to an identical mathematical objective. For single-hop queries, this approach transforms the complex multi-evidence problem to optimize search under a fixed top-$K$ constraint through progressive short-term memory construction. According to Eqs. \eqref{eq:1} and \eqref{eq:2}, the overall model is formulated as:
\begin{equation}
\begin{aligned}
    q(A, E | Q) 
    &= q(A, e_1, \dots, e_n | Q) \\
    &= q(A |E_1^{\text{new}} , \dots, E_J^{\text{new}}, Q) \prod_{j=1}^{J} q(E_j^{\text{new}} | q_j, m_j) \\
    &= f_{\theta}(E, Q) \prod_{j=1}^{J} f_\theta^{\text{memory}}(q_j, m_j, Q)
\end{aligned}
\end{equation}

Consequently, for multi-hop problems, according to Eqs. \eqref{eq:3} and \eqref{eq:4}, the sequential dependencies of the evidence chain are ultimately condensed into the exact same probabilistic structure:
\begin{equation}
\begin{aligned}
    q(A, E | Q)
    &= q(A, e_1, \dots, e_n | Q) \\
    &= q(A |e_{n} , \dots, e_1, Q) q(e_1, \dots, e_n| Q) \\
    &= f_{\theta}(E, Q) \prod_{j=1}^{J} f_\theta^{\text{memory}}(q_j, m_j, Q)
\end{aligned}
\end{equation}
Both paradigms are elegantly encapsulated into a single, cohesive equation:
\begin{equation}
\begin{aligned}
q(A, E | Q) &= f_{\theta}(E, Q) \prod_{j=1}^{J} f_\theta^{\text{memcot}}(q_j, m_j, Q)
\end{aligned}
\end{equation}

In summary, whether addressing the high-volume top-$K$ constraints of single-hop problems or navigating the sequential dependencies of multi-hop reasoning, our Multi-Step Memory architecture provides a mathematically unified modeling framework. Both paradigms are elegantly encapsulated into a single, cohesive equation, demonstrating the theoretical robustness and versatility of the proposed memory evolution mechanism.

\begin{table}[htbp]
\centering
\setlength{\tabcolsep}{3mm}
\caption{
The analysis of the cost and $F_1$ scores in the LoCoMo benchmark. Best results are highlighted in \textbf{bold}, and second-best results are \underline{underlined}.
}
\begin{tabular}{ll|lc}
\toprule
\textbf{Model} & \textbf{Method}
&\textbf{$F_1$ scores}&\textbf{Token}\\
\midrule

\multirow{7}{*}{\rotatebox{90}{\textbf{GPT-4o-mini\cite{4omini}}}}& Full Context&  35.57&16,910\\
 & RAG~\cite{rag}& 42.25&\textbf{649}\\
& Mem0~\cite{mem0}&  45.10&\underline{1,172}\\
& MemoryOS~\cite{li2025memosmemoryosai} &  42.84&1,589\\
& A-Mem~\cite{xu2025amem} &  39.65&2,520\\
& CompassMem~\cite{compassmem}&  \underline{52.18}&20,100\\
& \cellcolor[RGB]{235,245,250}MemCoT &  \cellcolor[RGB]{235,245,250}\textbf{58.03}&\cellcolor[RGB]{235,245,250}2,438\\
\midrule

\multirow{3}{*}{\rotatebox{90}{\textbf{Qwen2.5-14B\cite{qwen2025qwen25technicalreport}}}}& Full Context&  42.74&16,910\\
& RAG~\cite{rag}&  38.77&\textbf{695}\\
& Mem0~\cite{mem0} &  36.04&\underline{1,015}\\
& MemoryOS~\cite{li2025memosmemoryosai} &  40.28&1,400\\
& A-Mem~\cite{xu2025amem} &  28.98&1,137\\
&CompassMem~\cite{compassmem}&  \underline{52.52}&20,100\\
& \cellcolor[RGB]{235,245,250}MemCoT &  \cellcolor[RGB]{235,245,250}\textbf{57.06}&\cellcolor[RGB]{235,245,250}2,748\\
\bottomrule
\end{tabular}
\label{tab:tokens}
\end{table}
\begin{table}[htbp]
  \centering
\setlength{\tabcolsep}{1.8mm}
  \caption{The analysis of the average steps by category in the LoCoMo benchmark. The MH, TP, OD, and SH represent Multi-hop, Temporal, Open-domain, and Single-hop, respectively.}
  \label{tab:steps-mean-gpt4o-mini}
  \begin{tabular}{l|ccccc}
\toprule
    \textbf{Model} & \multicolumn{5}{c}{\textbf{Average Steps}}\\
    \cmidrule(lr){2-6}
 & \textbf{MH}&\textbf{TP} &\textbf{OD} &\textbf{SH} &\textbf{Overall}  \\
  \midrule
    GPT-4o-mini~\cite{4omini} & 1.83 & 2.41 & 3.25 & 1.52 & 1.87 \\
    Qwen2.5-7B~\cite{qwen2025qwen25technicalreport} & 1.90 & 2.69 & 2.55 & 1.76 & 2.03 \\
\bottomrule
  \end{tabular}
  \label{tab:costtokens}
\end{table}

\section{Additional Experiments}
\subsection{Computational Cost Analysis}
To study the cost of computation, we analyze the cost of tokens and the average evolution iteration $J$ of each category in the LoCoMo benchmark. As shown in Table~\ref{tab:tokens}, models of different sizes incur varying token costs. Notably, the maximum token consumption remains only 2,748, which is substantially lower than the full context length of 16,910 tokens. Although Mem0 or MemoryOS produce a relatively low cost of approximately 1.5k tokens, MemCoT explicitly makes a trade-off between the quality of the answer and the consumption of tokens. Compared to MemoryOS, MemCoT achieves a 41\% improvement in $F_1$ scores at the expense of an additional 1.3k tokens. Moreover, as shown in Table~\ref{tab:costtokens}, the average number of evolutionary iterations is merely 1.87 and 2.03 for GPT-4o-mini and Qwen2.5-7B, respectively. These results demonstrate that MemCoT effectively determines the optimal termination point for the iterative loop, even when the maximum number of iterations is set to $J=8$.

\begin{table*}[htbp]
\centering
\setlength{\tabcolsep}{3mm}
\caption{
The analysis of different types of retrieval methods in the LoCoMo benchmark. Best results are highlighted in \textbf{bold}. The model is GPT-4o-mini.
}
\begin{tabular}{ll|ccccc}
\toprule
  Type&\textbf{Method}
&    \textbf{Single-hop}&\textbf{Multi-hop}&\textbf{Open}&\textbf{Temporal}&\textbf{Overall}\\
\midrule

  &Full Context&      45.89&25.01&15.49&23.83&35.57\\
   &RAG~\cite{rag}&     52.19 &32.17 &23.21 &30.77 &42.25 \\
  Non graph&Mem0~\cite{mem0}&      47.65&38.72 &28.64 &48.93 &45.10 \\
  &A-Mem~\cite{xu2025amem} &      44.65 &27.02 &12.14 &45.85 &39.65 \\
  &\cellcolor[RGB]{235,245,250}MemCoT(w Na\"{\i}veRag~\cite{rag})&      \cellcolor[RGB]{235,245,250}\textbf{65.38}&    \cellcolor[RGB]{235,245,250}\textbf{43.85}&\cellcolor[RGB]{235,245,250}\textbf{32.99}&\cellcolor[RGB]{235,245,250}\textbf{56.10}&\cellcolor[RGB]{235,245,250}\textbf{57.73}\\
\midrule
  &MemoryOS~\cite{li2025memosmemoryosai} &      48.62 &35.27 &20.02 &41.15 &42.84 \\
 Graph& CompassMem~\cite{compassmem}&    57.36&38.84&26.61&57.96&52.18\\
  &\cellcolor[RGB]{235,245,250}MemCoT(w LightRag~\cite{guo-etal-2025-lightrag})&      \cellcolor[RGB]{235,245,250}\textbf{64.81} &\cellcolor[RGB]{235,245,250}\textbf{40.43} &\cellcolor[RGB]{235,245,250}\textbf{42.67} &\cellcolor[RGB]{235,245,250}\textbf{60.31} &\cellcolor[RGB]{235,245,250}\textbf{58.03} \\
\bottomrule
\end{tabular}
\label{tab:retrieval}
\end{table*}

\begin{table*}[t]
  \centering
  \caption{Performance on the LoCoMo benchmark with Advanced Models GPT-4.1-mini.Best results are highlighted in \textbf{bold}, and second-best results are \underline{underlined}.}
  \label{tab:high_cap_results}
  \resizebox{0.95\textwidth}{!}{%
  \begin{tabular}{l|l|cccccr}
  \toprule
  \textbf{Model}& \textbf{Method}& \textbf{Multi-hop}& \textbf{Temporal}& \textbf{Open-domain}& \textbf{Single-hop}& \textbf{Overall}& \multicolumn{1}{c}{\textbf{Token}} \\
  \midrule
  
  \multirow{8}{*}{\textbf{GPT-4.1-mini}} 
   & Full Context & 25.02 & 12.04 & 19.05 & 18.68 & 18.70  & 16,910 \\
   & MemoryBank~\cite{MemoryBank} & 5.00 & 5.94 & 5.16 & 5.72 & 5.46  & \textbf{432} \\
   & MemGPT~\cite{memgpt} & 17.72 & 19.44 & 11.29 & 25.59 & 18.51  & 16,977 \\
   & A-Mem~\cite{xu2025amem} & 25.06 & \underline{51.01} & 13.22 & 41.02 & 32.58  & 2,520 \\
   & Mem0~\cite{mem0} & 30.14 & 48.91 & 16.43 & 41.3 & 34.20  & 1,764 \\
   & Zep~\cite{zep} & 26.73 & 20.97 & 21.24 & 39.96 & 32.40  & \underline{1,602} \\
& MemCoT (w/o $\psi_{\text{z-o}}$)& \underline{38.51}& 38.94& \textbf{52.81}& \underline{64.58}&\underline{53.73}& 1,707\\
 \rowcolor{gray!10}& MemCoT& \textbf{43.89}& \textbf{60.88}& \underline{40.88}& \textbf{66.42}& 
\textbf{59.54}& 2,790\\
\bottomrule
  \end{tabular}
  }
  \label{tab:advanced_model}
  \end{table*}

\subsection{The Analysis of Retrieval Module}
Because of its short-term memory evolution mechanism, MemCoT is insensitive to the design of the retrieval method. To further analyze the role of the retrieval module in long-term memory, we conduct corresponding ablation studies. As shown in Table~\ref{tab:retrieval}, our method still maintains the best performance with both graph-based and non-graph-based retrieval methods, achieving $F_1$ scores of 58.03 and 57.73, respectively. It shows that the non-graph retrieval method will only slightly decline in the quality of the answer, while the graph retrieval method will get better performance at the cost of the time required for knowledge graph construction. Notably, we notice that no significant performance discrepancy between single-hop and multi-hop questions. We attribute this to the multi-step framework of MemCoT, which effectively bridges the gap between the non-graph and graph retrieval methods, which means the process of MemCoT builds a memory graph due to the short-term memory evolution module. Specifically, the short-term memory evolution module dynamically constructs an implicit memory graph during processing. For open-domain and temporal questions, graph-based methods such as LightRAG yield higher $F_1$ scores owing to their enhanced memory representations. Ultimately, these findings demonstrate that our method could be improved with the development of the retrieval methods.

\begin{figure*}[htbp]
    \centering
    \includegraphics[width=0.9\textwidth]{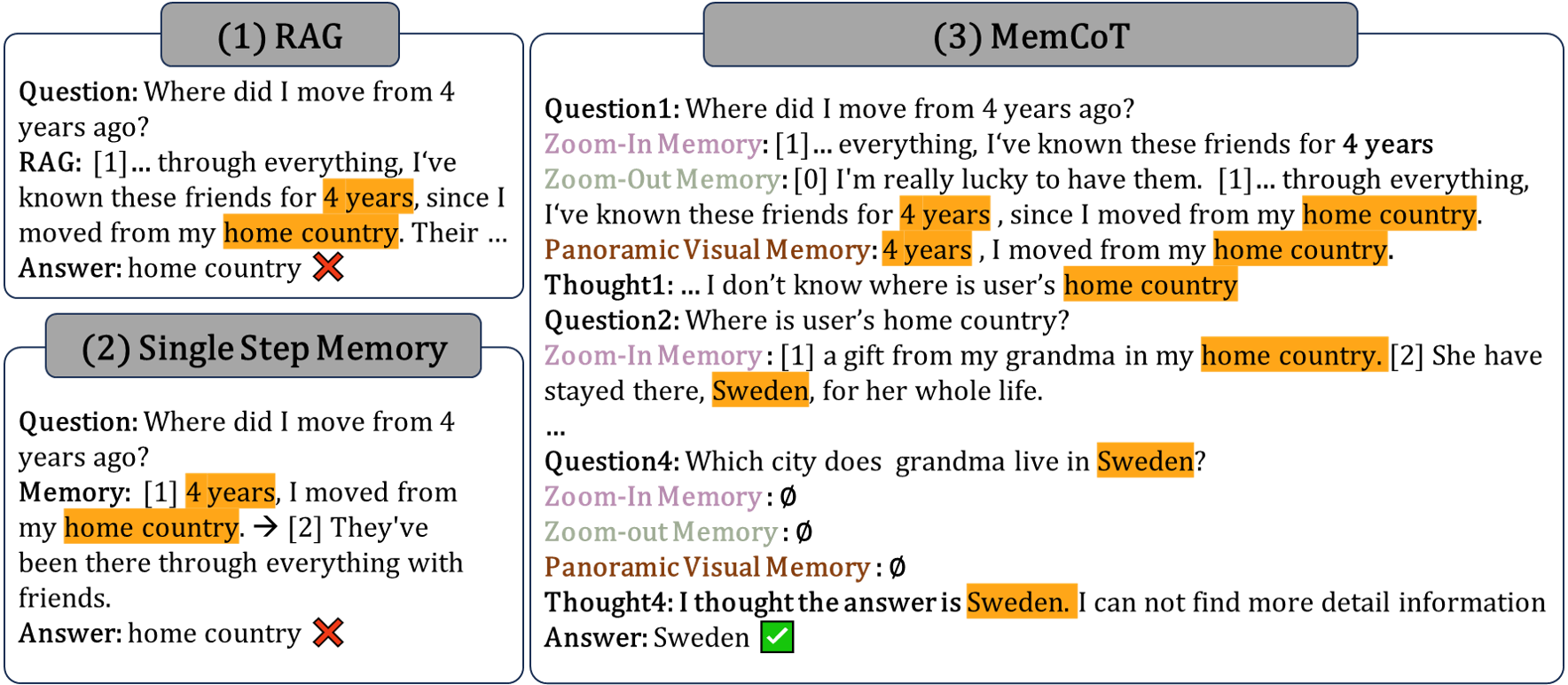}
    \caption{The specific comparison of different methods. The orange mark indicates the keywords of the truth label.}
    \label{fig:app_example}
\end{figure*}

\subsection{More Model Experiments}
To provide a more concrete illustration of MemCoT, Figure~\ref{fig:app_example} compares the workflows of standard RAG, the single-step memory baseline, and our proposed MemCoT. Specifically, MemCoT leverages diverse forms of short-term memory to facilitate the reasoning process, which in turn iteratively refines the search queries to retrieve more relevant memory. Furthermore, Table~\ref{tab:advanced_model} presents a detailed analysis of the trade-off between token consumption and response accuracy in the advanced model. Compared to Mem0, MemCoT achieves a substantial 83\% improvement in answer quality, although with a 58\% increase in token usage. Notably, when compared to Zep, the ablated variant MemCoT (w/o $\psi_{\text{z-o}}$) slightly increases only 6\% in the consumption of tokens while still delivering a 65\% improvement in quality, thereby demonstrating the robustness and high efficiency of the MemCoT framework.

\section{Detailed Settings}
For the long-term memory storage system, we adopt LightRAG as our underlying search engine. In both the LoCoMo and LongMemEval setups, we utilize gpt-4o-mini~\cite{4omini} for entity and relation extraction, aligning with standard practices in recent literature. To encourage diverse sampling, the generation temperature for agents within the multi-view long-term memory perception module is consistently set to 1.0. Conversely, the temperature for the judge agent is set to 0.0 to ensure deterministic and rigorous judgment on whether further query modifications are necessary. Finally, the responder agent's temperature is set to 1.0, which conforms to prevalent experimental configurations.
\\
\\
\section{Prompts.}
\lstset{
  language=Python,                  
  numbers=none,                     
  numberstyle=\small\color{gray},   
  stepnumber=1,                     
  breaklines=true,                  
  breakatwhitespace=false,          
  frame=single,                     
  tabsize=2,                        
  basicstyle=\ttfamily\small,       
  keywordstyle=\color{blue},        
  stringstyle=\color{orange},       
  commentstyle=\color{green!50!black}, 
  moredelim=**[is][\bfseries\color{red}]{@}{@}
}

To implement our results, we release the key prompts in our procedure. Below is the instruction for zoom-in focal retrieval.
\begin{lstlisting}
prompt = f"""
{query_information}
{known_information}
RAG retrieval results:
{rag_results_text}

Task: Identify useful retrieval chunks based on "Constraint Matching".

Analysis Rules:
1. **Temporal Check**: If the query mentions a specific time or duration, ANY result mentioning a matching time/duration is HIGHLY relevant.
2. **Descriptive Matches**: If the query asks for a specific name (e.g., a place or person), but the text only provides a generic description (e.g., "my place", "that person", "my origin"), mark it as USEFUL. It confirms the context.
3. **Partial Information**: Do not discard a result just because it lacks the final answer. If it provides the *background* or *cause* of the queried event, it is useful.

Output a JSON object:
1. thinking: A brief explanation covering, including why specific IDs were selected (mention the matching time or description).
3. missing_information: What specific missing_information is still missing. You can analyze the disadvantage of search_query.
4. useful_ids: List of indices (e.g., [1, 2]). Include IDs that contain matching timeframes or descriptions of the target, even if the specific name is missing.

Output ONLY valid JSON."""
\end{lstlisting}

Below is the instruction for zoom-out context expansion.
\begin{lstlisting}
prompt = f"""
{query_information}
{known_information}
{middle_context_text}

Task: Identify useful information to answer the query.

Analysis Rules:
1. **Temporal Check**: If the query mentions a specific time or duration, ANY result mentioning a matching time/duration is HIGHLY relevant.
1. **Descriptive Matches**: If the query asks for a specific name but the text only provides a generic description, mark it as USEFUL.
2. **Contextual Clues**: Surrounding turns may provide context, causation, or temporal references that help answer the query.

Output a JSON object:
1. thinking: Your reasoning about what useful information these context windows contain.
2. thinking_choice:  Why specific IDs were selected (mention the matching time or description); 
3. missing_information: What specific information is still missing to fully answer the query.
4. useful_ids: List of 0-based indices (e.g., [0, 2]) from the context windows that are useful.

Output ONLY valid JSON."""
\end{lstlisting}

Below is the instruction for panoramic visual grounding.
\begin{lstlisting}
prompt =f"""{query_information}
{known_information}
{rag_information}

You are viewing conversation session images from sessions: {session_list_str}.
Each image is a page from a conversation PDF. Dialogue entries are formatted as:
  {{dia_id}}- {{speaker}}: {{text}}
Some entries may include embedded images with captions.

Your task:
1. Carefully read ALL dialogue content visible in the images.
2. Identify dialogue entries (by their dia_id, e.g., "D1:5") that contain information relevant to answering the query.
3. Explain your reasoning.

Output a JSON object:
{{
"thinking": ...,
"useful_dia_ids": ...,
}}
Output ONLY valid JSON."""
\end{lstlisting}

Below is the instruction for the judge agent.
\begin{lstlisting}
prompt =f"""
Query: {query}
{short_memory_text}
{conv_memory_text}
{fail_queue_information}
Output a JSON object: 
1.{thinking}
2.useful_id: List of dia_id strings from the useful results (e.g., [0, 2]). If can_answer, include those that support the answer. If not, include those with relevant partial info.
3.can_answer: true if the results contain enough information to answer the query, false otherwise.
4. action: Check the **Fail query**. You can choose only one action to generate for each new query:
    1. Break: Break down the last query into sub-queries to get a shorter but more exact query. if Q=[Q_A,Q_B], you can just searcg Q_A firstly. Example: When Tom arrived at Shanghai 3 years ago-> [Tom arrived at Shanghai,3 years ago]
    2. Delete: If Root Query Q = [Q_A,Q_B] and Short Memory include Q_A, focus on Q_B and New query Q'=Q-Q_A.
    Do not let new_queries as same as and Fail query.
    You can try more types of action to avoid the same failed query.
5.new_queries: If can_answer is false, suggest {queries_num} new queries that are more likely to retrieve the missing information. These should be focused and based on the gaps identified in the report.
Output a JSON object exactly following this structure:
{{
"thinking": ...,
"useful_id": ...,
"can_answer": ...,
"action": ...,
"new_queries": ...,
}}
Output ONLY valid JSON."""
\end{lstlisting}

For the responder agent, we adopt the standard response prompts from LoCoMo~\cite{LOCOMO} and LongMemEval~\cite{longmemeval}, respectively.
\clearpage

\end{document}